\documentclass[preprint,aps,prc,amssymb,tightenlines,showpacs,endfloats]
{revtex4}
\usepackage{bm}                      % bold math
\usepackage[dvips]{graphicx}         % Include figure files

\begin{document}
\def\bea{\begin{eqnarray}}
\def\eea{\end{eqnarray}}
\def\be{\begin{equation}}
\def\ee{\end{equation}}
\def\rra{\right\rangle}
\def\lla{\left\langle}
\def\tv{\bm{\tau}}
\def\eps{\varepsilon}
\def\om{\omega}

\title{Temperature dependence of single-particle properties \\
       in nuclear matter}

%\date{\today}

\author{
W. Zuo,$^{1,2}$
\footnote{Corresponding address:
          Institute of Modern Physics, Chinese Academy of Sciences,
          P.O. Box 31, Lanzhou 730000, P.R. China.
          Tel: 0086-931-4969318; E-mail: zuowei@impcas.ac.cn}
Z. H. Li,$^{3}$ U. Lombardo,$^{4}$ G. C. Lu,$^{1,2}$ and H.-J.
Schulze$^{5}$}
\affiliation{
$^1$ Institute of Modern Physics, Chinese Academy of Sciences,
     Lanzhou 730000, P.R. China\\
$^2$ Graduate School of Chinese Academy of Sciences,
     Beijing 100039, P.R. China\\
$^3$ Institute of Theoretical Physics, Shanghai Jiao Tong University,
     Shanghai 200240, P.R. China\\
$^4$ INFN-LNS, Via Santa Sofia 44, I-95123 Catania, Italy\\
$^5$ Sezione INFN, Via Santa Sofia 64, I-95123 Catania, Italy}

\vskip2cm

\begin{abstract}
The single-nucleon potential in hot nuclear matter is investigated
in the framework of the Brueckner theory
by adopting the realistic Argonne $V_{18}$ or Nijmegen 93
two-body nucleon-nucleon interaction supplemented by a microscopic
three-body force.
%We particularly explore the temperature dependence
%of the three-body force effect on the single-particle potential.
The rearrangement contribution to the single-particle potential
induced by the ground state correlations is calculated in terms
of the hole-line expansion of the mass operator
and provides a significant repulsive contribution
in the low-momentum region around and below the Fermi surface.
Increasing temperature leads to a reduction of the effect,
while increasing density makes it become stronger.
The three-body force suppresses somewhat the ground
state correlations due to its strong short-range repulsion,
increasing with density.
Inclusion of the three-body force contribution results in a quite
different temperature dependence of the single-particle potential
at high enough densities as compared to that adopting the pure two-body force.
The effects of three-body force and ground state correlations
on the nucleon effective mass are also discussed.
\end{abstract}

%\vskip1cm

\maketitle

%------------------------------------------------------------------------------
\section{Introduction}

The determination of the equation of state (EOS) of nuclear matter
based on microscopic many-body approaches is of great interest in
nuclear physics and nuclear astrophysics
\cite{baldo:1998a,lombardo:2001a,lattimer:2000,danielewicz:2002}.
During the dynamical evolution of heavy ion collisions (HIC) at
intermediate and high energies, a transient state of hot and dense
nuclear matter can be produced and therefore the experiments with
HIC are powerful tools for constraining the nuclear EOS
\cite{danielewicz:2002}. Since the EOS can not be measured
directly in the experiments, one has to compare the experimental
observables and the theoretical simulations by using transport
models \cite{bertsch:1988}. The single-particle (s.p.) potential
felt by a nucleon in the nuclear medium is one of the basic
ingredients of transport models for HIC and controls together with
the nucleon-nucleon cross sections the collision dynamics.

Microscopically the s.p.~potential in cold nuclear matter has been
investigated by many authors based on microscopic theoretical
approaches \cite{baldo:1988a,baldo:1990,bombaci:1991,zuo:1999,sehn:1997}.
It has been pointed out \cite{brown:1971} that the s.p.~potential
calculated at the lowest level of the
Brueckner-Hartree-Fock (BHF) approximation cannot describe with
sufficient accuracy the mean field, but it is necessary to include
higher-order ground state correlations.
In Refs.~\cite{baldo:1988a,baldo:1988b}, the effect of the
ground-state correlations on the s.p.~potential has been
investigated within the Brueckner-Bethe-Goldstone (BGG) theory.
Therein it is shown that inclusion of the rearrangement
contribution in the s.p.~potential is also crucial for restoring
the Hugenholtz-Van Hove theorem, which is strongly violated
at the lowest BHF level of approximation.
The importance of the rearrangement term in the hole-line expansion
of the mass operator has also been verified in
Refs.~\cite{jeukenne:1976,zuo:1998,lombardo:2001b} in connection
with the optical model potential, the s.p.~properties in the nuclear
medium such as the nucleon effective mass and the nucleon mean free
path, and the superfluidity properties in neutron matter and
nuclear matter.
Recently, the calculation of the high-order
correlations in the mass operator has been extended to the case of
isospin asymmetric nuclear matter at zero
temperature \cite{zuo:1999} within the extended BHF approach.

Nowadays it is widely recognized that dressing in a
non-relativistic framework the interaction with short-range
correlations, either G-matrix \cite{baldo:1998a}
or in-medium T-matrix \cite{schnell,Czerski:2002,dewulf:2003,frick},
is not enough to reproduce the empirical saturation properties
of cold nuclear matter.
The saturation mechanism demands for the
high-density repulsive contribution of the three-body force (TBF)
\cite{lejeune:1986,grange:1989}.
Microscopic TBF \cite{grange:1989} have been used in Brueckner
calculations \cite{zuo:2002a,zhou:2004}.
Phenomenological versions of TBF have also been used mainly in
variational approaches \cite{panda}.
In Ref.~\cite{zuo:2004} we have extended
the BHF approach with microscopic TBF to finite temperature
and investigated the EOS of hot nuclear matter.
It was shown that the TBF affect considerably the properties
of hot nuclear matter especially at high densities and temperatures.
The aim of the present paper is to extend our
previous work by including the effect of the ground state
correlations in the calculation of the single-nucleon properties.

The paper is arranged as follows.
In the next section~II we shall
give a brief review of our theoretical model including the
finite-temperature BHF approach with a microscopic TBF
and the hole-line expansion for the mass operator at finite
temperature \cite{jeukenne:1976,mahaux:1991}.
Our numerical results are presented and discussed in Sect.~III,
including the rearrangement term and its temperature dependence,
the TBF effect on the ground state correlations and the s.p.~potential.
Finally, a summary will be given in Sect.~IV
along with a short comparison with other approaches.

%------------------------------------------------------------------------------
\section{Theoretical Model}

\subsection{Finite-temperature Brueckner-Hartree-Fock approach with
a three-body force}

The general formalism of the Brueckner-Bethe-Goldstone (BBG)
theory for cold nuclear matter can be found in
Refs.~\cite{bombaci:1991,zuo:1999} and its extension to the finite
temperature case is given in Ref.~\cite{bombaci:1994}.
Here we give a brief review for completeness.
The starting point of the BBG scheme is the Brueckner reaction G-matrix,
which satisfies the following Bethe-Goldstone (BG) equation,
\be
 G[\rho,T,\om] = V + V \sum_{k_1k_2} \frac{
 |k_1k_2\rangle Q(k_1,k_2,\rho,T) \langle k_1k_2|}
 {\om-\eps(k_1)-\eps(k_2)+i0}
 G[\rho,T,\om] \:,
\label{e:bg}
\ee
where $\om$ is the starting energy and $V$ denotes the
realistic nucleon-nucleon (NN) interaction.
The finite-temperature Pauli operator $Q$ can be expressed as
\be
 Q(k_1,k_2,\rho,T) = \left[1-f(k_1)\right] \left[1-f(k_2)\right] \:,
\label{e:q}
\ee
where $f(k)$ is the Fermi distribution at finite temperature,
\be
 f(k) = \left[ 1+\exp\left(\frac{\eps(k)-\mu}{T}\right)\right]^{-1} \:.
\label{e:f}
\ee
In terms of the normalization condition,
\be
 \rho = \sum_{k}f(k) \:,
\ee
one can determine the chemical potential $\mu$ self-consistently by
iteration for any given density $\rho$ and temperature $T$.
The s.p.~energy $\eps(k)$ in
Eqs.~(\ref{e:bg}) and (\ref{e:f}) is defined as
\be
 \eps(k) \equiv \eps(k,\rho,T) = \frac{k^2}{2m} + U(k,\rho,T) \:,
\label{e:e}
\ee
where the s.p.~potential $U(k,\rho,T)$ may be calculated from
the real part of the on-shell anti-symmetrized G-matrix,
\be
 U(k,\rho,T) = \sum_{k'} f(k')\,
 {\rm Re} \left\langle kk' \right|
 G[\rho,T,\eps(k)+\eps(k')] \left| kk' \right\rangle_A \:.
\label{e:u}
\ee
In the present calculations, the continuous
choice \cite{jeukenne:1976} is adopted for the s.p.~potential.
On the one hand, in the zero temperature limit it provides a much
faster convergence of the hole-line expansion than the gap
choice \cite{song:1998};
on the other hand, it appears a natural choice for $T\ne 0$,
since in the finite-temperature case any
distinction between particles and holes becomes meaningless.
Within the continuous choice the s.p.~potential describes at the BHF
level the mean field felt by a nucleon during its propagation
between two successive scatterings in the nuclear
medium \cite{lejeune:1978}.

The NN interaction $V$ in the present calculation contains two parts:
the Argonne $V_{18}$ \cite{wiringa:1995} or the
%Argonne $V_{14}$ \cite{wiringa:1984}
Nijmegen 93 \cite{nij93}
two-body interaction plus the contribution of a microscopic TBF.
Two kinds of TBF have been adopted in the BHF formalism:
One is the semi-phenomenological Urbana TBF \cite{pudliner:1995},
which has two or few adjustable parameters determined by fitting the empirical
saturation density and energy of cold symmetric nuclear matter
in the BHF calculations \cite{baldo:1997,baldo:1999,zhou:2004}.
The other one (used here)
is a microscopic TBF based on meson exchange coupled to
intermediate virtual excitations of nucleon-antinucleon pairs and
nucleon resonances, which was originally proposed in \cite{grange:1989}.
It contains the contribution of the two-meson
exchange part of the NN interaction medium-modified by the
intermediate virtual excitation of nucleon resonances,
the term associated to the non-linear meson-nucleon coupling required by
chiral symmetry,
the simplest contribution arising from meson-meson interactions,
and finally the two-meson exchange diagram with the virtual excitations
of nucleon-antinucleon pairs.
In this TBF model, the four important mesons $\pi$, $\rho$, $\sigma$, and
$\om$ are considered \cite{machleidt:1989}.
The parameters of the TBF, i.e., the coupling constants and the form factors,
have been redetermined recently in Ref.~\cite{zuo:2002a} from the
one-boson-exchange potential (OBEP) model to meet the
self-consistent requirement with the adopted two-body force.
A more detailed description of the TBF model and the approximations can
be found in Ref.~\cite{grange:1989}.

In the zero-temperature case, the TBF contribution $W_3$ has been
included in the BHF calculations by constructing an effective
two-body interaction $\overline{V_3}$
via a suitable average with respect to the
third-nucleon degrees of freedom \cite{grange:1989,lejeune:1986}.
By extending this scheme to finite temperature,
one can reduce the TBF to a temperature-dependent effective
two-body force $\overline{V_3}(T)$, which reads in
$r$-space \cite{zuo:2004}
\bea
 \langle \bm r_1' \bm r_2'| \overline{V_3}(T)| \bm r_1 \bm r_2 \rangle
 &=&
 \frac{1}{4} {\rm Tr} \sum_{k_n} f(k_n)
 \int d{\bm r_3} d{\bm r_3'} \,
 \phi^*_n(r_3') \left[1-\eta(r_{13}',T)\right] [1-\eta(r_{23}',T)]
\nonumber\\[0mm] && \times\,
 W_3(\bm r_1' \bm r_2' \bm r_3' | \bm r_1 \bm r_2 \bm r_3)
 \phi_n(r_3) [1-\eta(r_{13},T)] [1-\eta(r_{23},T)] \:,
\label{e:tbf}
\eea
where the trace is taken with respect to the spin and isospin of
the third nucleon.
The defect function $\eta(r,T)$
\cite{grange:1989,lejeune:1986} is defined as
$\eta(r,T) = \phi(r) - \psi(r,T)$,
where $\psi(r,T)$ is the correlated wave function for the relative
motion of two nucleons in the nuclear medium and $\phi(r)$ is the
corresponding unperturbed one.
A detailed description and justification of the above scheme
can be found in Ref.~\cite{grange:1989}.
As has been pointed out in Refs.~\cite{zuo:2004,zuo:2004a},
the TBF $W_3$ itself is the same as the one adopted in our previous
calculations for the zero-temperature case \cite{zuo:2002a}
and is independent of temperature.
However, in the finite-temperature case, the
effective two-body force $\overline{V_3}(T)$
depends on temperature due to the medium effects
caused by the Fermi distribution $f(k)$
and the defect function $\eta(r,T)$, which is
strongly temperature dependent.
%As a consequence the contribution of the TBF
%is expected to be more pronounced at finite temperature.

In the BHF approximation with the TBF,
Eqs.~(\ref{e:bg}) and (\ref{e:e}-\ref{e:tbf})
are solved self-consistently.
First, the temperature-dependent G-matrix is calculated along with the
auxiliary potential by solving the BG equation, then one
evaluates the defect function with only the two-body force, further
constructs the effective two-body force, and finally adds the
effective two-body force to the bare two-body force.
This procedure is repeated until convergence is reached.
Obviously, the effect of microscopic TBF is
automatically included into the G-matrix by iteration.

\subsection{Hole-line expansion of the mass operator}

Within the BBG theory the mass operator can be expanded into a
perturbation series in terms of the number of hole
lines \cite{day:1978,jeukenne:1976,grange:1987}:
\be
 M(k,\om) = M_1(k,\om) + M_2(k,\om) + M_3(k,\om) + \ldots \:,
\ee
which is schematically illustrated in Fig.~\ref{f:hl}.
In this expansion the two-hole line contribution
$M_2(k,\om)$ is the so-called rearrangement term,
representing the Pauli rearrangement,
i.e., the medium dependence of the effective
G-matrix interaction via the ground state Pauli blocking effect.
The third-order term $M_3(k,\om)$
accounts for the fact that the hole state is partially empty in
the correlated ground state of nuclear
matter \cite{jeukenne:1976,baldo:1990}.

The mass operator
$M(k,\om )=V(k,\om)+iW(k,\om)$ is a complex quantity.
When evaluated on the energy shell, its real part
$V(k) \equiv V(k,\eps(k))$
describes the s.p.~potential felt by a nucleon
in the nuclear medium and can be compared with the empirical potential
depth extracted from the optical potential model \cite{jaminon:1989},
whereas its imaginary part
$W(k) \equiv W(k,\eps(k))$ is related to the nucleon mean free
path \cite{jeukenne:1976,zuo:1998,zuo:1999}.
The on-shell condition is given by the following energy-momentum
relation \cite{zuo:1999},
\be
 \frac{k^2}{2m} + {\rm Re}\, M(k,\om) = \om \:.
\label{e:on}
\ee
To the lowest order of approximation, the on-shell condition is
simplified as
$k^2\!/2m+{\rm Re}\, M(k,\eps(k))=\eps(k)$,
where $\eps(k)$ is the BHF s.p.~energy.
In the present work, we shall consider only the on-shell properties of
the mass operator.
Hereafter we denote the real and imaginary
parts of the on-shell mass operator as $V(k)$ and $W(k)$, respectively.
Their hole-line expansions can be written as
\bea
 V(k) &=& V_1(k)+V_2(k)+V_3(k)+\ldots \:,
\\
 W(k) &=& W_1(k)+W_2(k)+W_3(k)+\ldots \:.
\eea

Since the renormalization contribution $M_3(k)$ is quite small
compared to the lowest-order BHF term $M_1(k)$ and the
rearrangement term $M_2(k)$ \cite{grange:1987,zuo:1999},
in the present paper we concentrate on the investigation of the
rearrangement contribution $M_2(k)$,
which in the finite-temperature case can be expressed as follows,
\be
 M_2(k_1,\om) = \frac{1}{2} \sum_{k_2 k'_1 k'_2}
 [1-f(k_2)] f(k'_1) f(k'_2)\,
 \frac{ |\langle k_1 k_2 | G | k'_1k'_2 \rangle_A|^2 }
 {\om + \eps(k_1)-\eps(k'_1)-\eps(k'_2)+i0} \:.
\ee
By using the usual angular averaging procedure
in order to remove the angular
dependence of the energy denominator and the anti-Pauli operator
\cite{brueckner:1958,baldo:1991,baldo:1998a}, $M_2(k,\om)$ can be
readily calculated in terms of the G-matrix expanded in the
partial wave representation \cite{zuo:1999,grange:1987}:
\bea
 M_2(k,\om) &=& \frac{2}{\pi^2 k} \sum_{JSTLL'}(2J+1)
 \int q dq \int P dP \left[ 1 - f(\sqrt{P^2\!/2+2q^2-k^2}) \right]
\nonumber \\  && \times
 \int q'^2 dq' \,R(q',P)\,
 \frac{ \big| G_{LL'}^{JST}[q,q',P,\eps_{12}(q',P)] \big|^2 }
 {\om + \eps(\sqrt{P^2\!/2+2q^2-k^2}) - \eps_{12}(q',P) + i0} \:,
\eea
where
$\bm{q'}=(\bm{k}_1-\bm{k}_2)/2$ and
$\bm{P}=\bm{k}_1+\bm{k}_2$
are the relative and total momenta, respectively.
In the above equation,
$\eps_{12}(q',P) \equiv \langle \eps(k_1) + \eps(k_2) \rangle$
denotes the angular average of the energy denominator
and the angular-averaged anti-Pauli operator is defined as
$R(q',P) \equiv \frac{1}{2} \int_0^\pi
\sin\theta d\theta f(k_1)f(k_2)$,
where $\theta$ is the angle between $\bm{q}'$ and $\bm{P}$.

Thus in terms of the G-matrix obtained from Eq.~(\ref{e:bg}), the first- and
second-order contributions in the hole-line expansion of the mass
operator can be calculated.
Hereafter we shall denote the on-shell mass operator containing the
contribution of the rearrangement term as
$M_{12}(k)=M_1(k)+M_2(k)$,
and its real and imaginary parts as
$V_{12}(k)=V_{1}(k)+V_{2}(k)$ and
$W_{12}(k)=W_{1}(k)+W_{2}(k)$, respectively.

%------------------------------------------------------------------------------
\section{Results and Discussion}

The self-consistent BHF procedure extended to TBF has been
applied to study symmetric nuclear matter at finite temperature.
Two realistic interactions,
the Argonne $V_{18}$ and Nijmegen 93 potentials,
have been used to describe the two-body force,
whereas the model of TBF is the meson exchange interaction discussed
in Sec.~IIA.
The results for the EOS have been presented elsewhere \cite{zuo:2002a}.
Here we only report the saturation properties:
The main effect of TBF is
to reduce the saturation density from 0.26 fm$^{-3}$ to 0.19 fm$^{-3}$
due to the high-density extra repulsion.
The energy per particle rises from $-18$ MeV to $-15$ MeV,
whereas the compression modulus is lowered from 230 MeV to 210 MeV.

\subsection{Rearrangement contribution}

In Fig.~\ref{f:vw} are plotted the real and imaginary parts of the
rearrangement contribution $M_2(k)$.
It is seen that they depend sensitively on temperature in
both cases with and without including the TBF contribution.
The real part of $M_2(k)$, i.e., the rearrangement term $V_2(k)$ of
the s.p.~potential is repulsive and its contribution is mainly
concentrated in the region below the Fermi momentum $k_F$, where
the ground state hole-hole correlations are expected to be most
effective.
Around $k_F$ the magnitude of $V_2(k)$ decreases
rapidly as a function of momentum and vanishes at high enough momentum.
As the nuclear matter is heated up, $V_2(k)$ is
strongly reduced due to the softening of the Pauli blocking around
the Fermi surface at high temperature,
which weakens the effect of the ground state hole-hole correlations.

Also the TBF contribution suppresses considerably these
correlations in cold and hot nuclear matter due to its strong
short-range repulsion, and leads to a sizable reduction of the
rearrangement correction $V_2(k)$. This feature is much more
pronounced in the low-momentum region, where the ground state
correlations are more significant. The influence of the TBF
diminishes as the temperature increases, since the ground state
correlations themselves become smaller at higher temperature. For
instance, in the case of the $V_{18}$ potential, at $T=0$, the
reduction of $V_2(k=0)$ due to the TBF is about 6 MeV, from 30 MeV
to 24 MeV, while at $T=20$ MeV, the reduction is about 3 MeV, from
13 to 10 MeV. Comparing the curves in the upper panels with the
corresponding ones in the lower panels, it is readily seen that in
both cases with the $V_{18}$ potential and with the Nijmegen 93
potential, the influence of the TBF decreases as the temperature
rises, implying that the temperature dependence of the TBF effect
on the correlation potential $V_2(k)$ is not much sensitive to the
adopted two-body realistic NN interactions. However, the effect of
the TBF is more pronounced in the case with the Nijmegen 93
potential in the temperature range considered here.

The imaginary part $W_2(k)$ of $M_2(k)$ is related directly to the
lifetime or the width of a hole state \cite{jeukenne:1976}, which
vanishes above the Fermi momentum $k_F$ for cold nuclear matter
due to the Pauli blocking effect \cite{zuo:1999}.
However, at finite temperature, the tail of $W_2$
slightly extends to the momentum region above the Fermi surface,
since the Fermi surface becomes diffusive and
the Pauli blocking is weakened in this case.
The temperature dependence of $W_2(k)$ is somewhat
more complicated than that of the real part $V_2(k)$.
With increasing temperature, $W_2(k)$ gets smaller in the momentum
region well below $k_F$, while it becomes larger in the upper part
of the Fermi sea.
This can be explained as follows.
On the one hand, the ground state hole-hole correlations decrease with
temperature.
On the other hand, at a higher temperature, a hole
state intends to decay faster,
especially close to the Fermi surface.
The competition between these two effects determines the
final variation of $W_2(k)$ as a function of temperature.
The TBF suppresses somewhat the ground state hole-hole correlations
and its contribution reduces the magnitude of the imaginary part
$W_2(k)$ in agreement with the results obtained for the real part
of $M_2(k)$.

In order to discuss the density dependence, we report in Fig.~\ref{f:m}
the rearrangement contribution $M_2$ at the fixed momentum $k=0$
as a function of density for the two cases with and without the TBF.
It is seen that the real and the imaginary parts of $M_2(k)$
increase monotonically with density in both cases,
indicating that the effect of the ground state hole-hole
correlations is stronger in denser nuclear matter,
where the number of hole state is larger.
The TBF leads to a reduction of
the ground state correlations in the whole density region considered.
Its effect is fairly small at low densities and
becomes stronger rapidly as the density increases.

\subsection{Mass operator}

The complex mass operator $M_{12}(k)$ including the BHF
contribution and the rearrangement term is reported in Fig.~\ref{f:mi}
for several different values of density
and two temperatures $T=0$, 20 MeV.
Since the real part of the rearrangement term is repulsive,
its contribution reduces to a large extent the attraction of the
pure BHF s.p.~potential $V_1(k)$ \cite{zuo:2004}.
In the zero-temperature case
at normal nuclear matter density $\rho=0.17$ fm$^{-3}$,
the repulsive contribution of the ground
state correlations causes the depth of the s.p.~potential
$V_{12}(k=0)$ to rise to $\approx -60$ MeV
from its BHF value $V_1(k=0) \approx -85$ MeV \cite{zuo:1999}.
This improves considerably the
agreement of our predicted s.p.~potential with the optical model
potential extracted from nucleon-nucleus scattering
experiments \cite{jeukenne:1976,jaminon:1989}.
As the nuclear density increases, the total potential $V_{12}(k)$ becomes more
attractive in the low-momentum region, while at high enough
momenta it becomes slightly less attractive.
Such a density dependence is mainly attributed to the density behavior of the
pure BHF s.p.~potential $V_1(k)$,
as in the zero-temperature case \cite{lombardo:2001a}.

Concerning the temperature dependence,
it is seen from Fig.~\ref{f:mi} that in the case without TBF,
the total potential becomes more attractive at low momenta and less
attractive in the high momentum region above $k_F$
when the nuclear matter is heated up.
This is quite different from the
temperature behavior of the lowest-order BHF s.p.~potential,
where an increase in temperature results
in an overall reduction of the attraction of the BHF s.p.~potential
in the whole momentum region \cite{zuo:2004}.
At low momenta, the
enhancement of the attraction of $V_{12}(k)$ with temperature is
readily understood in terms of the temperature effect on the
ground state correlations:
The correlation potential $V_2(k)$
applies mainly to states below the Fermi momentum and its
repulsive contribution is reduced largely by increasing
temperature as shown in Fig.~\ref{f:vw}.
At high momenta, the contribution
of $V_2(k)$ is very small and consequently the variation of the
total potential $V_{12}(k)$ with temperature is essentially
determined by that of the BHF potential $V_1(k)$.

At low densities (for example, $\rho=0.085$ fm$^{-3}$),
the TBF effect is fairly weak.
However, at high densities, the TBF
modification of the s.p.~potential becomes significant,
especially for high temperatures.
In the two cases of $\rho=0.225$ and 0.34 fm$^{-3}$,
inclusion of the TBF contribution even makes the
temperature behavior of $V_{12}(k)$ quite different from that without TBF.
It can been seen from the figure that
the calculated $V_{12}$ with the TBF contribution gets more
repulsive in the whole momentum region.
This may be understood as
a consequence of the competition between the following two effects.
Below the Fermi momentum, the TBF
suppresses the ground state hole-hole correlations due to its
strong short-range repulsion and thus it reduces the repulsive
contribution of the rearrangement term $V_2(k)$.
On the other hand, the TBF contribution to the BHF s.p.~potential $V_1(k)$ is
repulsive and reduces the attraction of $V_1(k)$.
At high densities, as the temperature increases, the TBF effect on the
ground state correlations becomes weaker while its contribution to
the pure BHF s.p.~potential $V_1(k)$ gets larger.
Above the Fermi momentum, the correlation term $V_2(k)$ tends to vanish and the
temperature dependence of the total potential $V_{12}$ is
dominated by the BHF one $V_1(k)$.
Accordingly at high densities
the total potential gets more repulsive in the whole momentum
region as the matter is heated.
Another feature related to the
temperature dependence that can be observed from Fig.~\ref{f:mi}
is that the
curvature around the Fermi momentum becomes more smooth with
increasing temperature in both cases with and without TBF,
which is in agreement with the prediction obtained for
the pure BHF s.p.~potential \cite{lejeune:1986,zuo:2004} and is
attributed to the thermal excitations around the Fermi surface at
finite temperature.

Now let us turn to the imaginary part $W_{12}$, which is also
called absorptive potential \cite{jeukenne:1976}.
It is seen from Fig.~\ref{f:mi}
that for cold nuclear matter the absorptive potential
crosses zero at the Fermi momentum $k_F$.
The BHF contribution $W_1(k)$ vanishes below $k_F$ and the
correlation term $W_2(k)$ above $k_F$ due to the Pauli blocking.
At finite temperature, $W_1(k)$ may extend to the momentum region below
$k_F$ and $W_2(k)$ to above $k_F$.
The total absorptive potential
crosses zero at a momentum close to $k_F$, since the chemical
potential does not change very much up to $T=20$ MeV \cite{grange:1987}.
For each fixed density considered here,
the absolute value of $W_{12}(k)$ increases above the Fermi
momentum $k_F$ while it decreases below $k_F$.
The temperature
dependence of $W_{12}(k)$ turns out to be stronger in nuclear
matter with a smaller density.
As expected, the TBF effect on the
imaginary part $W_{12}(k)$ increases with density, while it is
relatively weak compared to that on the real part $V_{12}(k)$,
especially at high densities.

It is instructive to make a comparison with the predictions
of the in-medium T-matrix approach, where the self-energy is
self-consistently calculated with the T-matrix
\cite{schnell,Czerski:2002,dewulf:2003,frick}.
The main difference
from the G-matrix is that the T-matrix embodies both particle-particle and
hole-hole correlations.
Since its convergence is not protected by any hole line expansion,
one needs a very accurate determination of the self-energy.
Also, zero-temperature calculations are usually not possible in this approach
without removing in some way the too strong pairing instabilities.
Doing so, however, nuclear matter appears to be underbound
\cite{Czerski:2002,dewulf:2003,frick}.
Concerning the s.p.~properties,
one observes an overall qualitative agreement between the
T-matrix self-energy and the Brueckner one
(see for instance Fig.~5 of Ref.~\cite{Czerski:2002}
or Fig.~13 of Ref.~\cite{dewulf:2003}),
which is an indication that the
rearrangement term contains most of the hole-hole correlations.

\subsection{Effective mass}

The effective mass describes the nonlocality of the s.p.~potential
and makes its local part less attractive.
It is of great interest \cite{lunney:2003}, since it is closely related with
many nuclear phenomena such as the dynamics of heavy ion
collisions at intermediate and high energies \cite{cugnon:1987},
the damping of nuclear excitations and giant resonances \cite{bertsch:1983},
and the adiabatic temperature of collapsing stellar matter \cite{onsi:2002}.
The effective mass $m^*(k)$ is defined as \cite{jeukenne:1976}
\be
 \frac{m^*(k)}{m} = \frac{k}{m} \left[\frac{dE(k)}{dk}\right]^{-1} \:,
\ee
where $E(k)$ is the s.p.~energy determined by the momentum-energy
relation Eq.~(\ref{e:on}).
When the mass operator is expanded up to the second order,
one can readily calculate the effective mass as follows:
\be
 \frac{m^*(k)}{m} =
 \left[ 1 + \frac{m}{k}\frac{dV_{12}(k)}{dk} \right]^{-1} \:.
\ee

The calculated $m^*(k)/m$ versus momentum is reported in Fig.~\ref{f:ms},
where the upper panel shows the results for
nuclear matter at the empirical saturation density $\rho=0.17$ fm$^{-3}$
and for two temperatures $T=0$, 20 MeV with and without including TBF,
while the lower panel displays the results for three values of density
$\rho=0.085$, 0.17, and 0.34 fm$^{-3}$ at zero temperature.
Due to the high possibility for particle-hole excitations near the Fermi
surface \cite{jeukenne:1976}, the momentum dependence of
$m^*(k)/m$ in cold nuclear matter is characterized by a wide bump
around the Fermi momentum $k_F$.
Recently such a structure has
also been found \cite{dalen:2005} within the relativistic
Dirac-Brueckner-Hartree-Fock approach for the momentum dependence
of the non-relativistic type of effective mass introduced in terms
of the Schr\"odinger equivalent s.p.~potential \cite{jaminon:1989}.
Inclusion of the contribution of
the rearrangement term makes the peak of the effective mass more
pronounced as compared to the results obtained at
the lowest BHF level of approximation \cite{zuo:2004},
consistent with previous investigations adopting pure two-body NN
interactions \cite{jeukenne:1976,grange:1987,baldo:1988a,zuo:1999}.
The value $m^*_F \equiv m^*(k_F)$ of the effective mass
obtained in the present calculation is around $1.02$ when the TBF
contribution is included and $1.08$ in the case without the TBF.
Both values are larger than the BHF value $\approx 0.8$ \cite{zuo:1999},
which is mainly attributed to the
contribution of the ground state correlations, i.e., the
arrangement term $V_2(k)$.
Inclusion of even higher-order terms,
i.e., the third- and the fourth-order terms in the hole-line
expansion of the mass operator may reduce $m_F^*/m$ to about 0.9
as discussed in Ref.~\cite{zuo:1999}.

As the nuclear matter is heated, the peak of $m^*(k)$ becomes
flatter and the peak value lower, which is
related directly to the temperature effect on the s.p.~potential
$V_{12}$ around the Fermi momentum and is similar to the result
obtained at the BHF level of approximation \cite{zuo:2004}.
It is seen that in the case of $T=20$ MeV
the peak structure almost disappears.
The TBF effect on the
effective mass is significant only at low temperatures and in the
low-momentum region below $k_F$, because the TBF effect on the BHF
s.p.~potential $V_1(k)$ is an overall enhancement in the whole
momentum region \cite{zuo:2004}
and thus the momentum-dependence of the total potential $V_{12}(k)$
is mainly affected by the correlation potential $V_2(k)$, which is
important below $k_F$.
As the temperature increases, the
correlation potential $V_2(k)$ itself and the influence of the TBF
on it become weaker as has been shown in Fig.~\ref{f:vw},
and as a result also the TBF effect on the effective mass is smaller
at higher temperature.
From the lower part of Fig.~\ref{f:ms} it is clear
that with increasing density
the peak structure of the effective mass around $k_F$ becomes
less pronounced and the value of the effective mass at $k_F$ decreases,
which is comparable with the
previous calculations adopting pure two-body nucleon-nucleon
forces \cite{baldo:1988a,zuo:1998}.

%------------------------------------------------------------------------------
\section{Summary}

In the present work, we have reported the investigation of the
s.p.~properties for cold as well as hot nuclear matter within the
framework of the Brueckner theory by adopting the Argonne $V_{18}$
or the Nijmegen 93 two-body nucleon-nucleon interaction
plus a microscopic three-body force based on the meson-exchange model.
The mass operator has been calculated up to the second order of the
hole-line expansion.
Special attention has been paid to the effect
of ground state correlations on the s.p.~potential in hot nuclear
matter and the influence of the TBF on the ground state correlations.
Our result shows that these
correlations give a repulsive contribution to the s.p.~potential
mainly within the Fermi sphere in agreement with previous
investigations \cite{jeukenne:1976,zuo:1999}.
As the temperature rises,
the real and the imaginary part of the rearrangement
contribution become less repulsive due to the weakening of the
correlations.

The TBF contribution turns out to
reduce the two-hole line correlation term in magnitude due to its
strong short-range repulsion.
When the nuclear matter is heated,
the TBF effect on the ground state correlations becomes weaker.
At $T=0$, the imaginary part of the second-order mass
operator vanishes below the Fermi surface due to the Pauli
blocking, while in the finite-temperature case its tail may
extend slightly above the Fermi momentum, since the
Fermi surface becomes diffusive and the Pauli blocking is weakened.
Both the real and the imaginary parts of the
rearrangement term are shown to be increasing functions of density,
indicating that the effect of the ground state
correlations is stronger at higher density.

The consideration of the two-hole line diagram of the mass
operator reduces remarkably the attraction of the lowest-order BHF
s.p.~potential and improves significantly the agreement of our
microscopic s.p.~potential with the empirical optical model
potential \cite{jeukenne:1976}.
For low density up to about $\rho_0=0.17$ fm$^{-3}$,
the total s.p.~potential $V_{12}(k)$ is shown to become more
attractive in the low-momentum region
when the nuclear matter is heated up,
which can be attributed to the reduction of the repulsive contribution
from the ground state correlations in the finite-temperature case.
The role played by the TBF turns out to become stronger as the
density increases.
The TBF affects the total s.p.~potential in two ways.
On the one hand it provides a repulsion to the BHF part
and makes the BHF s.p.~potential more repulsive.
On the other hand it suppresses
somewhat the ground state correlations and reduces the repulsive
contribution of the rearrangement term.
As a combined result, at
high densities the s.p.~potential becomes less attractive in the
whole momentum region when the TBF is taken into account.
At high enough densities, the TBF contribution even
changes the temperature dependence of the s.p.~potential, i.e., it
makes the s.p.~potential more repulsive at high temperature than
at low temperature.

In both cases with and without the TBF, the momentum dependence of
the effective mass displays a broad peak around the Fermi momentum
in the zero-temperature case, which is similar to the result
obtained in the lowest BHF approximation.
The effects of the TBF
and the ground state correlations on the effective mass are found
to be important mainly below the Fermi surface.
The ground state correlations make the peak more pronounced
as compared to the BHF one,
whereas the TBF reduces the peak via its effect on the rearrangement term.
Increasing temperature smoothes the s.p.~potential around
the Fermi surface and consequently leads to a flattening of the
peak, making it almost vanish at high enough temperature.
The value of the effective mass at the Fermi
surface decreases with density, as in the BHF approximation.

In summary, it is shown by our investigation that both the ground
state correlations and the TBF affect considerably the s.p.~potential
in the nuclear medium and its temperature dependence.
It is therefore important to take into account their effects in the
application to transport-model simulations of HIC.

This investigation reinforces the role played by TBF in nuclear
matter. The G-matrix, as two-body effective interaction, which
embodies the short-range particle-particle correlations, while
preventing hard-core divergences, is not able to reproduce the
empirical saturation density at the convergence of the hole line
expansion \cite{song:1998}. Other non-relativistic approaches, the
variational method \cite{panda} as well as in-medium T-matrix
theory seem to confirm the latter conclusion. On the other hand,
it has been proven that the Dirac-Brueckner theory includes TBF
(Z-diagrams) as the effect of coupling to the negative energy
states.

\section*{Acknowledgments}

This work has been supported in part by the National Natural
Science Foundation of China (10575119,10235030), the Knowledge
Innovation Project (KJCX2-SW-N02) of CAS, the Major Program for
Basic Research Development of China (G2000077400), the Major
Prophase Research Project of Fundamental Research of the Ministry
of Science and Technology (2002CCB00200) of China, and by the
Asia-Link project (CN/ASIA-LINK/008(94791)) of the European
Commission.

%--------|---------|---------|---------|---------|---------|---------|---------|
%\newpage
%\baselineskip 0.28in

%--------|---------|---------|---------|---------|---------|---------|---------|

\newpage
\begin{figure}[h]
\caption{
Hole-line expansion of the mass operator.}
\label{f:hl}
\end{figure}

%\newpage
%\includegraphics[height=12.5cm,angle=0,bb=20 20 20 230]{tdspp2.ps.gz}
%\includegraphics[height=1.cm,angle=0,bb=320 0 320 550]{Fig2.ps.gz}
%\includegraphics[height=1.cm,angle=0]{Fig2.ps.gz}
%\vskip170mm
\begin{figure}[h]
\caption{
The real part $V_2(k)$ and the imaginary part $W_2(k)$ of
the rearrangement term $M_2$ versus nucleon momentum $k$
at normal nuclear density $\rho_0=0.17$ fm$^{-3}$.
The results in the upper and lower panels are
obtained by adopting the $V_{18}$ and Nijmegen 93
two-body nucleon-nucleon interactions, respectively.
The solid (dashed) curves
represent the results with (without) the TBF contribution
for temperatures $T=0,10,20$ MeV from top to bottom.}
\label{f:vw}
\end{figure}

%\newpage
\begin{figure}[h]
\caption{
The density dependence of the real and the imaginary parts of the
rearrangement term at $k=0$, $M_2(k=0)$,
with (solid lines) and without (dashed lines) the TBF contribution
for three different temperatures $T=0, 10, 20$ MeV from the top to bottom.}
\label{f:m}
\end{figure}

%\newpage
\begin{figure}[h]
\caption{
The real and imaginary parts of the total mass operator
up to the second-order term
for four different nuclear densities
as a function of momentum $k$.
In each panel the rising curves correspond to the real part $V_{12}(k)$
and the dropping curves to the imaginary part $W_{12}(k)$.
For both quantities, the solid and dashed (dotted and dot-dashed)
lines stand for the results at $T=0,20$ MeV
with (without) the TBF contribution,
as indicated in the first panel.}
\label{f:mi}
\end{figure}

%\newpage
\begin{figure}[h]
\caption{
The effective mass $m^*\!/m$ including the BHF and the second-order
correlation contributions as a function of momentum $k$.
The upper panel shows the results at normal nuclear matter density
for two temperatures $T=0, 20$ MeV with and without the TBF.
The lower panel depicts the effective mass
including the TBF contribution
at $T=0$ for three densities
$\rho =$ 0.085, 0.17, and 0.34 fm$^{-3}$.}
\label{f:ms}
\end{figure}

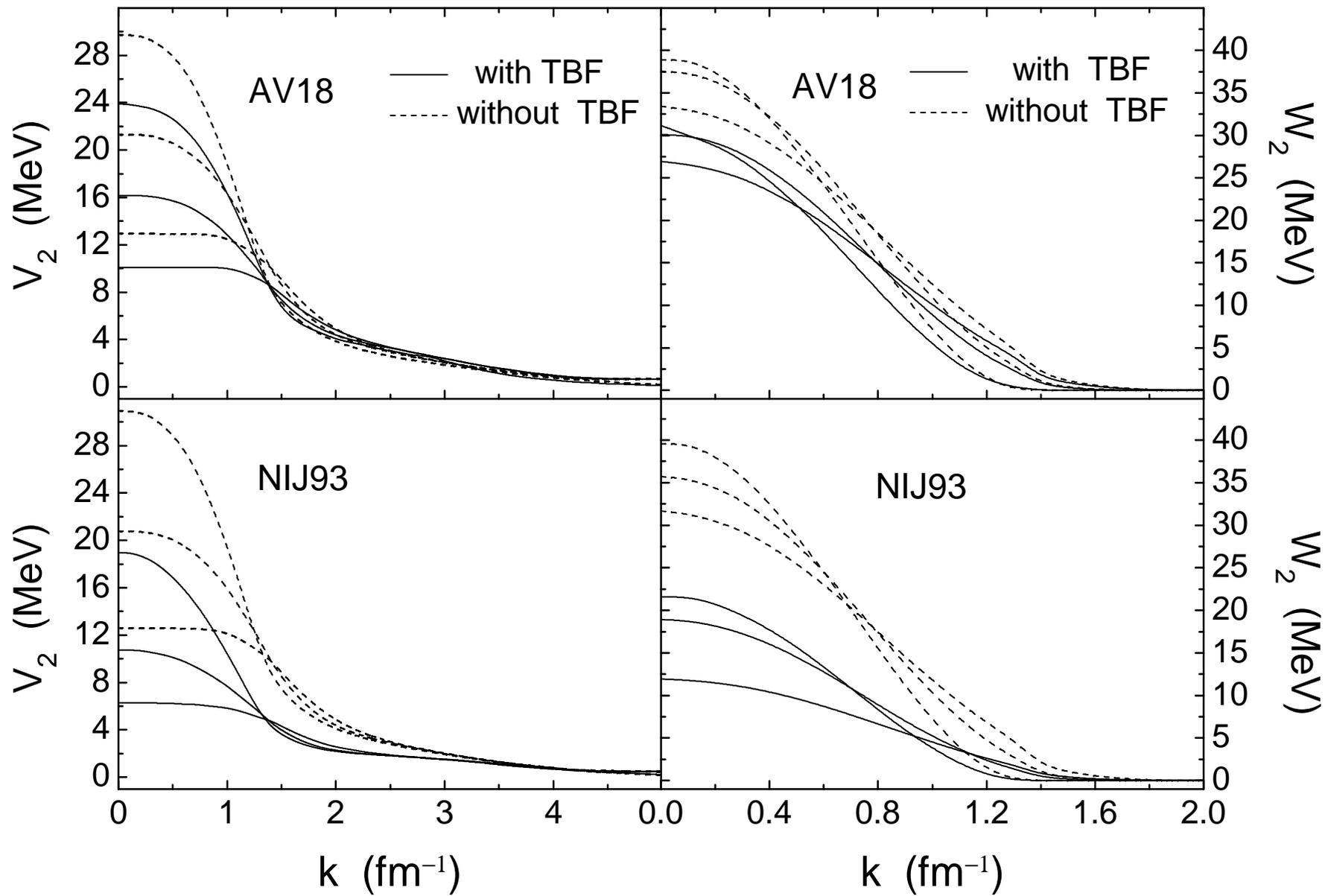
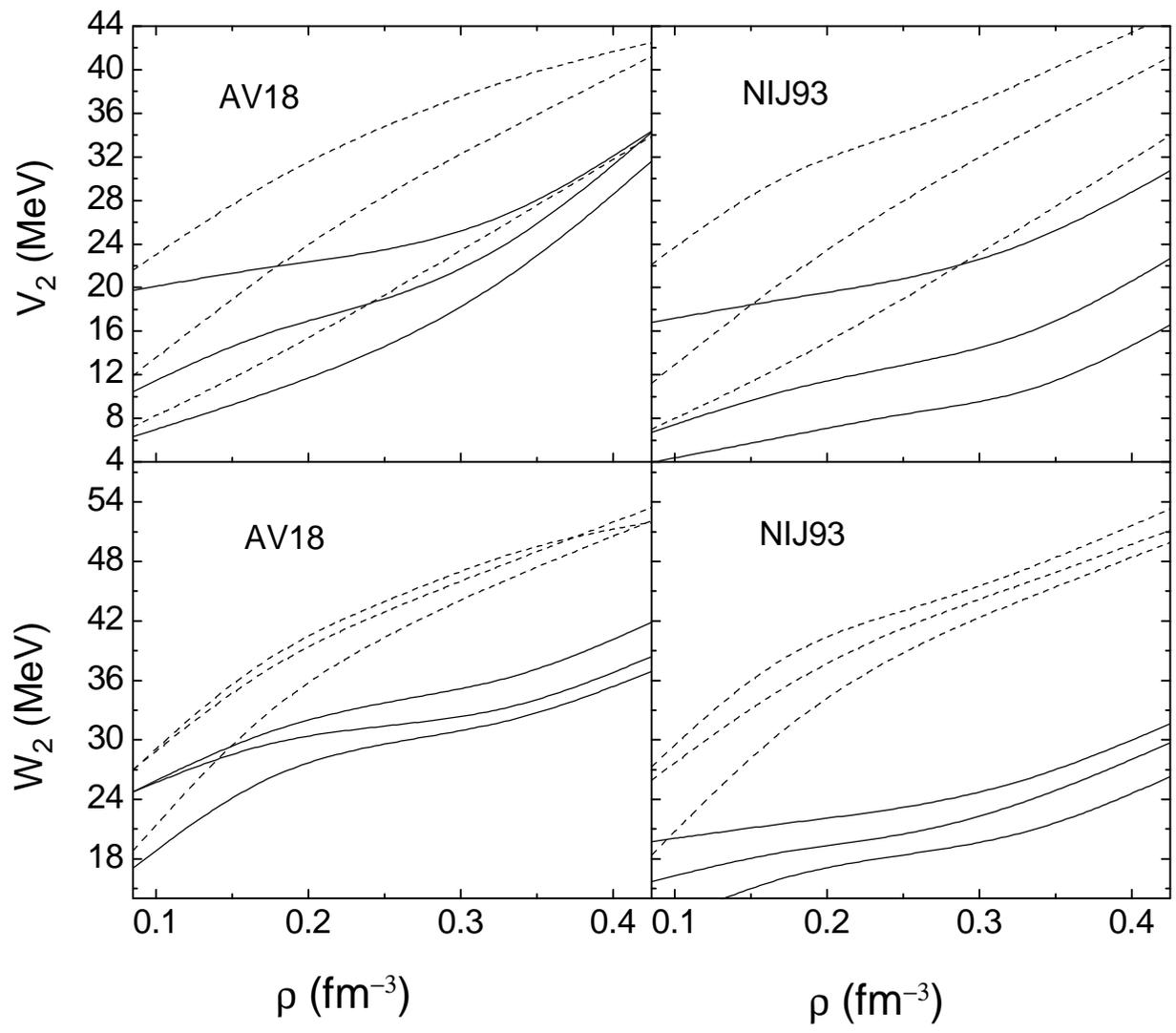
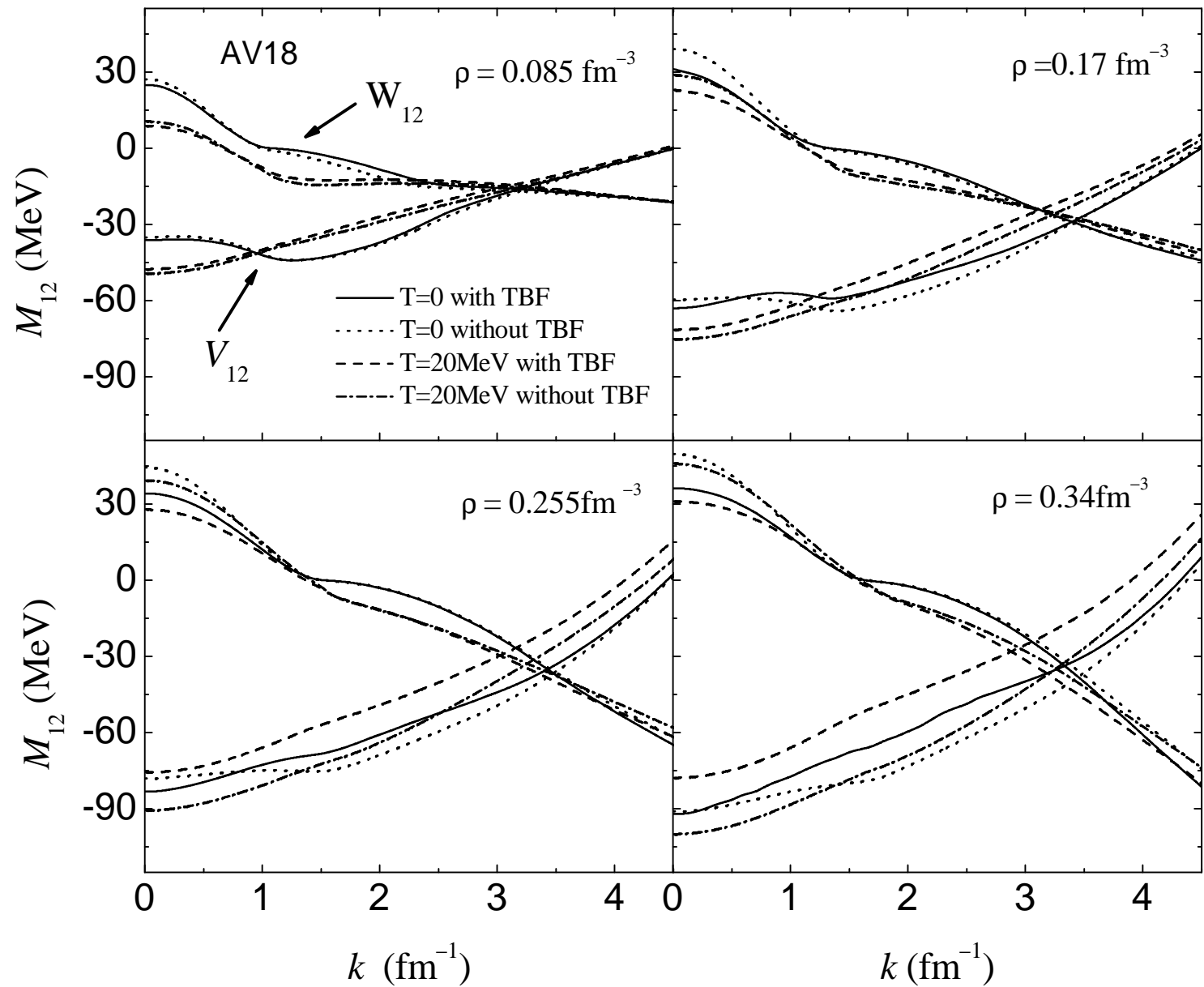
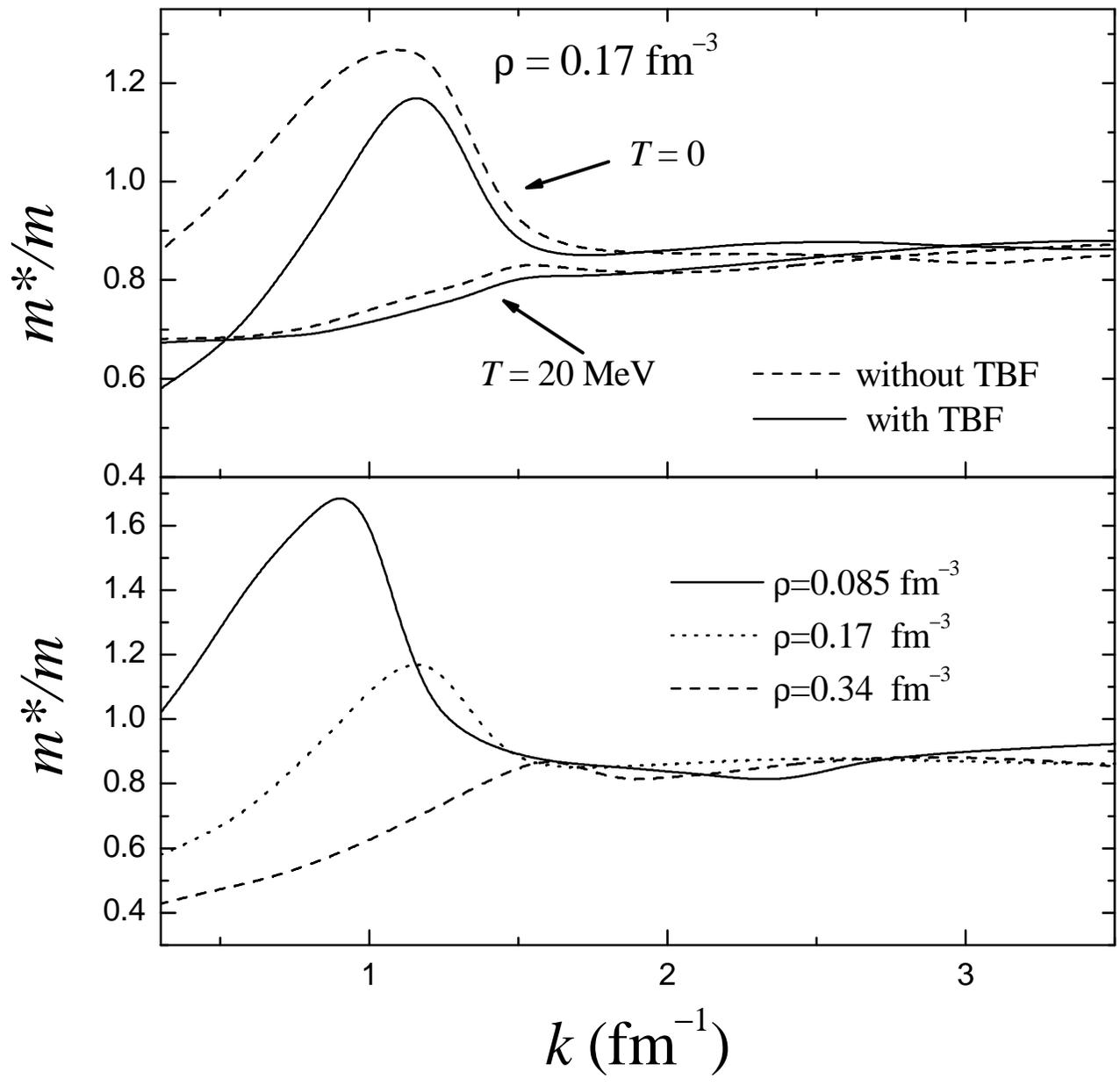
\end{document}